# Investigating and understanding the effects of multiple femtosecond laser scans on the surface topography of metallic specimens

Edwin Jee Yang **Ling**[a] (edwin.ling@mail.mcgill.ca)

Julien **Saïd**[a] (julien.said@ecl2015.ec-lyon.fr)

Nicolas **Brodusch**[b] (nicolas.brodusch@mcgill.ca)

Raynald **Gauvin**[b] (raynald.gauvin@mcgill.ca)

Phillip **Servio**[a] (phillip.servio@mcgill.ca)

Anne-Marie **Kietzig**[a,c] (anne.kietzig@mcgill.ca)

a. Department of Chemical Engineering, McGill University, 3610 University Street, Montréal, Québec, H3A 0C5, Canada.

b. Department of Mining and Materials Engineering, McGill University, 3610 University Street, Montréal, Québec, H3A 0C5, Canada.

c. Corresponding author. Tel: +1(514) 398-3302.

**Bolded** words indicate the authors' family name.

**Abstract**

The majority of studies performed on the formation of surface features by femtosecond laser radiation focuses on single scan procedures, i.e. solely manipulating the laser beam once over the target area to fabricate different surface topographies. In this work, the effect of scanning stainless steel 304 multiple times with femtosecond laser pulses is thoroughly investigated over a wide range of fluences. The resultant laser-induced surface topographies can be categorized into two different regimes. In the low fluence regime ($F_{\Sigma line,max} < 130$ J/cm$^2$), ellipsoidal cones (randomly distributed surface protrusions covered by several layers of nanoparticles) are formed. Based on chemical, crystallographic, and topographical analyses, we conclude that these ellipsoidal cones are composed of unablated steel whose conical geometry offers a significant degree of fluence reduction (35-52%). Therefore, the rest of the irradiated area is preferentially ablated at a higher rate than the ellipsoidal cones. The second, or high fluence regime ($F_{\Sigma line,max} > 130$ J/cm$^2$) consists of laser-induced surface patterns such as columnar and chaotic structures. Here, the surface topography showed little to no change even when the target was scanned repeatedly. This is in stark contrast to the ellipsoidal cones in the first regime, which evolve and grow continuously as more laser passes are applied.

**Keywords**

1. Femtosecond laser ablation
2. Irradiated metals
3. Laser treatment
4. Surface
5. Formation mechanism

## 1. Introduction

Femtosecond (fs) laser surface processing is a high-precision technique used to impart micro- and nano-sized features to a material surface. The resulting features can be broadly categorized into laser-inscribed and laser-induced structures [1]. Laser-inscribed structures consist of machined features such as grooves and holes, whose dimensions are equal to or greater than the effective laser beam diameter [2, 3]. On the other hand, laser-induced surface structures that are formed under laser irradiation have feature sizes smaller than the effective beam diameter, and the latter has been the subject of extensive research in recent years. Laser-induced structures that have been discovered on metals include laser-induced periodic surface structures (LIPSS) [4-6], bumps/cones [7-22], holes [7, 9, 17, 18, 23], undulating grooves [7, 9, 17], melt-like [18, 24], and chaotic structures [7, 9, 25]. These laser-induced structures can be used to alter the wettability of a material, resulting in a more hydrophobic or hydrophilic surface [10, 11, 26]. Other uses of such laser-induced topographies include surface coloration [27, 28], surface-enhanced Raman scattering (SERS) [29], reduced cell growth [30, 31], and improved broadband optical absorption [15, 32] for applications such as photovoltaics [33]. A comprehensive review of fs-laser surface texturing on metals has been provided by Ahmmed *et al*. [1].

In their recent work, Ahmmed *et al*. [7] investigated the surface topographies formed on titanium, stainless steel, aluminum and copper by fs-laser irradiation over a wide range of processing parameters. They observed that when an irradiated area is re-scanned multiple times at a particular experimental setting, the surface features become increasingly regular and well-defined. Similarly, Noh *et al*. [13] fabricated highly regular arrays of bumps and pillars on NAK80 mold steel by repeating their laser scan thirty and fifty times, respectively. Kam *et al*. [11] observed the formation of two distinct types of micro-cones on stainless steel 316L, which only appeared after the target was scanned more than one hundred times. Finally, studies by Zuhlke *et al*. [18-21] have expounded on the dominant formation mechanisms of these laser-induced surface structures, notably via stop-motion scanning electron microscopy (SEM) imaging. Despite the numerous reports on the fabrication of surface features by employing multiple laser scans, it is still unclear how re-scanning a previously-irradiated target influences the final surface topography.

Therefore, the present work investigates the effect of multiple laser scans on the resultant surface structures on stainless steel 304 over a large range of fluences (32 J/cm$^2$ < $F_{\Sigma line,max}$ < 1096 J/cm$^2$) and number of scans (up to 30). We report the changes in surface topography that occur during repeated scanning, which, in turn, provide an improved understanding of the underlying processes that govern the formation of micro- and nano-scale surface features on laser-irradiated metals.

**2. Experimental**

*2.1. fs-laser processing*

We used an amplified Ti:Sapphire laser (Coherent Libra) with a wavelength of 800 nm, a repetition rate of 10 kHz, and a pulse duration of <100 fs to produce laser-induced surface features on stainless steel 304 (0.030" thickness, McMaster-Carr) samples polished to a roughness of $R_a$ = 43 nm. These samples were manipulated under the horizontally-polarized Gaussian laser beam by a linear *x-z* translation stage (Zaber Technologies, Inc.) in a raster scan pattern. The target was maintained at a distance of Δ*y*

= +1.5 mm from the focal point, meaning that the machining plane was located between the focal point and the 100 mm focusing lens; this yielded a theoretical $1/e^2$ beam diameter $\omega_o$ of 58 µm. A variable attenuator composed of a half-wave plate and a polarizing beam splitter reduced the output power of 4 W to the desired processing power $P$. The raster scanning speed was set at $v$ = 4 mm/s, which corresponds to a horizontal pulse displacement of $\Delta x$ = 0.4 µm. Since the width of ablated lines ($\omega_{eff}$) laser-etched onto a target varies depending on the experimental conditions ($P$, $v$, $\Delta y$, sample material and gaseous environment), the vertical displacement $\Delta z$ was fixed to $\varphi_{line}$ = 80% of $\omega_{eff}$ for each combination of laser processing parameters, where $\varphi_{line}$ is the vertical line overlap. The number of laser scans, or passes applied to each patch is denoted by $N_s$. After fs-laser micromachining, the steel samples were cleaned with acetone in an ultrasonic bath for five minutes to remove any residual nanoparticle debris.

*2.2. Surface Analysis*

*2.2.1. SEM and EDS*

The laser-induced surface topographies were analyzed and imaged using SEM (FEI Inspect F50). Stop-motion SEM imaging (Section 3.1), first introduced by Zuhlke *et al*. [18], was performed as follows. Once the fs-laser had scanned the target in a raster pattern, the latter was cleaned with acetone in an ultrasonic bath for five minutes and then imaged with SEM. The sample was then returned to the translational stage and carefully re-aligned to the same starting position as before, where the second laser pass was subsequently applied ($N_s$ = 2). The sample was cleaned again and imaged with the SEM at the same location and magnification. This process was repeated for each additional laser scan applied, resulting in a series of SEM images that tracked the growth of ellipsoidal cones with each scan. Energy dispersive X-ray spectroscopy (EDS) on metallic specimens (Section 3.2) was conducted using the same SEM, which was equipped with a 60 mm$^2$ silicon drift detector (Octane Super, EDAX®), and analyzed using the TEAM$^{TM}$ EDS Analysis System.

*2.2.2. ECCI and EBSD*

In order to obtain crystallographic information pertaining to the material within the ellipsoidal cones (Section 3.3), the cones were sectioned along the transverse and longitudinal planes as follows. For the transverse section, we filled the irradiated area containing ellipsoidal cones with superglue to prevent them from fracturing. After drying, the specimen was grinded and polished using standard metallographic procedures until the cones were abraded to approximately half of their original height. The last polishing step was performed using a colloidal suspension of 50 nm silica particles. For the longitudinal section, we cut the laser-irradiated steel sample using a high-precision diamond saw through the middle of the target patch. After careful mechanical polishing (identical to the procedure followed for the transverse section), the specimen cross-section was milled in a flat ion milling system (Hitachi IM3000) with Ar$^+$ ions for several hours, using an accelerating voltage of 6 kV, a ion probe current of 80 µA and an angle of incidence of 80º.

Electron channeling contrast imaging (ECCI) micrographs were recorded with a retractable solid-state photo-diode backscattered electron detector (PD-BSE) attached to a Hitachi SU-8000 cold-field emission scanning electron microscope (CFE-SEM) (Hitachi High-Technologies Canada Inc.). A beam energy of 5 kV and a probe current of approximately 1 nA were used for imaging, and the beam convergence angle at the specimen surface was approximately 12.5 mrad. The electron backscatter diffraction (EBSD) analysis was conducted with the same CFE-SEM with an accelerating voltage of 20 kV and 4-5 nA probe current. The EBSD system from HKL (Oxford Instruments US) consisted of a Nordlys II EBSD camera controlled by the Flamenco software from the Channel 5 package. A 2×2 camera binning acquired the EBSD patterns, which resulted in 672×512 pixels² pattern images. Integration of two frames with 30 ms dwell time yielded the raw pattern images, and up to 8 Kikuchi bands were used for the orientation calculations. The crystallographic phase used for indexing was iron fcc (space group 225, F m -3 m) with a lattice parameter of 3.66 Å. The EBSD maps were post-processed using Tango from the same software package.

*2.2.3. 3D confocal microscopy*

The heights and diameters of ellipsoidal cones were measured using 3D confocal microscopy (Olympus LEXT OLS4000) (Section 3.4). Since each laser-irradiated patch contained several ellipsoidal cones, we determined $h$ and $d$ for multiple cones. Furthermore, each ellipsoidal cone was measured using two orthogonal baselines as their base area was elliptical.

*2.3. Model computations*

The four independent processing parameters ($P$, $v$, $\Delta y$, $\varphi_{line}$) can be condensed into one variable, $F_{\Sigma line,max}$, using the accumulated fluence profile ($AFP$) model, first developed by Eichstädt *et al.* [34]. Using $F_{\Sigma line,max}$ to report the processing parameters is preferred over the conventional use of peak fluence ($F = 8P/\pi\omega_o^2$) because it fully captures all the experimental settings. This irradiation model calculates the total deposited laser energy on the substrate by summing the fluence distribution of Gaussian pulses in both horizontal and vertical displacements, i.e. $\Delta x$ and $\Delta z$ respectively. The derivation and other details regarding the *AFP* model can be found in Refs. [7, 12, 34]. All model computations were executed using *MATLAB®*.

## 3. Results and Discussion

Figure 1 displays three representative laser-induced surface patterns obtained on stainless steel 304 and their evolution with increasing $N_s$. Multiple scan experiments at $F_{\Sigma line,max}$ = 32 J/cm² (Figure 1a) revealed the existence of isolated bumps distributed randomly over a knobby and lumpy terrain covered by LIPSS. At $N_s$ = 1, the laser-irradiated steel surface exhibits a LIPSS topography decorated with a few small and randomly distributed surface protuberances. As $N_s$ is increased, larger bumps begin to appear over the underlying LIPSS terrain. These bumps increase in diameter and eventually merge with adjacent ones as $N_s$ is increased even further. A close-up SEM image of a representative bump obtained at $F_{\Sigma line,max}$ = 39 J/cm² and $N_s$ = 10 is shown in Figure 2.

Here we observe a singular bump, approximately 34 μm in diameter and 20 μm in height, which possesses sloped sides and a rounded top. Since its longitudinal cross-section resembles an ellipse, we refer to it as an "ellipsoidal cone". When the stainless steel 304 sample is ablated by fs-laser pulses, nanoparticles are ejected from the surface, which then deposit and adsorb onto the sample surface. With the exception of its topmost section, the ellipsoidal cone is entirely covered by spherical nanoparticles on the order of 50-100 nm in diameter that have agglomerated into larger clumps. The individual nanoparticles remain distinguishable from one another, which leads to a fragile arrangement of globular nanoparticles.

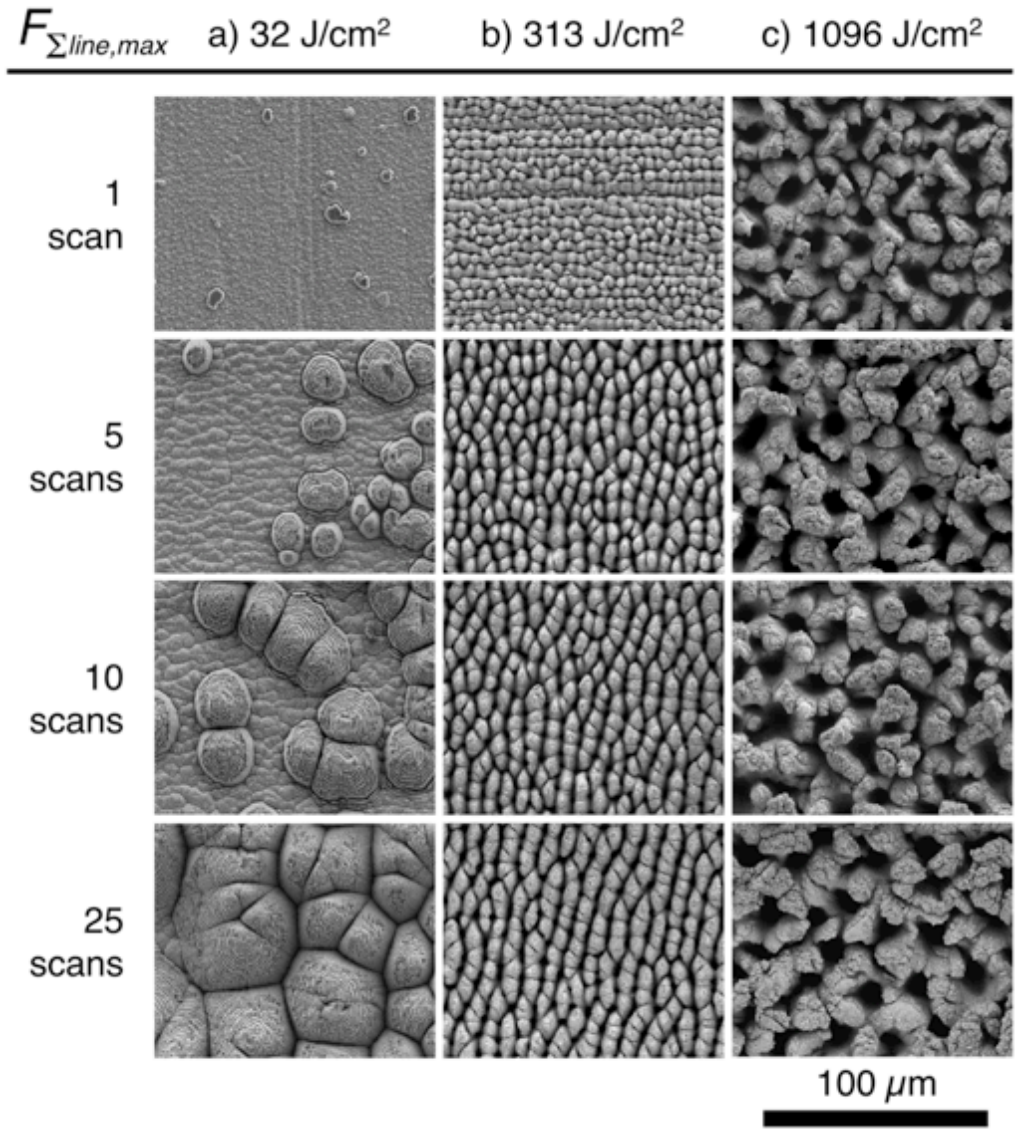

**Figure 1.** Evolution of fs-laser-induced surface topographies with increasing $N_s$ for three different values of $F_{\Sigma line,max}$. a) Ellipsoidal cones, b) columnar structures, and c) chaotic structures.

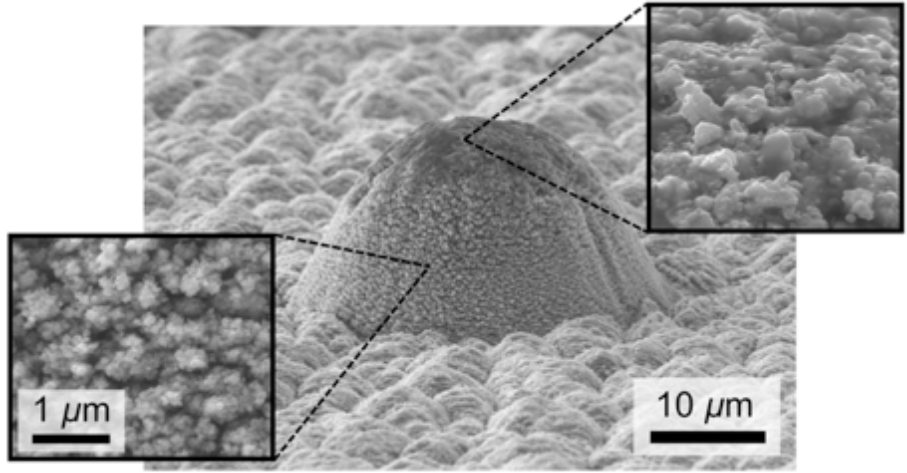

**Figure 2.** A representative ellipsoidal cone obtained at $F_{\Sigma line,max}$ = 39 J/cm$^2$ and $N_s$ = 10, viewed from a tilt angle of 75º. The 1 µm scale bar applies to both insets.

The top region of the ellipsoidal cone in Figure 2, on the other hand, is covered by a different melt-like morphology, where the nanoparticles appear to be sintered together by fs-laser irradiation. Cleaning of the stainless steel sample by ultra-sonication in acetone occasionally provokes the fracture and partial removal of the nanoparticle layer, revealing the presence of underlying strata, as demonstrated in Figure 1a. An interesting observation from Figure 1a is that the number of strata on the sides of the ellipsoidal cones corresponds to the number of laser scans applied to the target (as also highlighted in Figure S1 of the accompanying Supplementary Data).

The surface features obtained at $F_{\Sigma line,max}$ = 313 J/cm$^2$ (Figure 1b) are well-documented in literature as columnar structures [1], while those shown in Figure 1c at $F_{\Sigma line,max}$ = 1096 J/cm$^2$ are chaotic structures that show no repeated pattern [7, 9]. By visual inspection of Figure 1b and 1c, it is evident that, for both columnar and chaotic structures, increasing $N_s$ has little effect on the resultant surface topography. The only exception is for $F_{\Sigma line,max}$ = 313 J/cm$^2$: at $N_s$ = 1, the surface pattern consists of "undeveloped" and linked bumps that are unevenly arranged in horizontal tiers, which bear more resemblance to the undulating grooves described in [7]; at $N_s$ = 5, on the other hand, the separation between adjacent bumps becomes more distinguishable, with the bumps being larger, more regular, and more aligned; this was also observed in [7]. From $N_s$ = 5 onwards, the surface topography no longer varies even though $N_s$ is increased. This shows that, for columnar and chaotic structures, the laser-induced surface pattern obtained at $N_s$ = 1 ultimately governs the surface topography that arises for higher values of $N_s$.

Columnar (Figure 1b) and chaotic (Figure 1c) structures are obtained at higher fluence levels than ellipsoidal cones, and their formation is thus coupled with complex factors such as plasma plume formation and expansion. Zuhlke *et al*. [18] have proposed two formation mechanisms for the similar surface topographies they observed on nickel based on preferential valley ablation and hydrodynamics, but a comprehensive ablation model encompassing all the complexities of high fluence ablation is still lacking in literature. We leave the latter discussion for future work. In this study, we investigate the formation and growth of ellipsoidal cones (Figure 1a and 2) using a series of chemical, crystallographic, and topographical analyses. While several authors [11, 19, 20, 35-40] have already observed the formation of ellipsoidal cones via laser processing, this work focuses on elucidating the mechanisms behind the formation and growth of ellipsoidal cones by multiple laser raster scans.

*3.1. Stop-motion SEM imaging*

One of the peculiarities of the ellipsoidal cones is that they only appeared after the target area was re-scanned multiple times. Figure 3 displays selected frames from a stop-motion SEM sequence for a specimen irradiated at $F_{\Sigma line,max} = 47$ J/cm$^2$ to provide a perception of the evolution of the size and shape of the ellipsoidal cones as $N_s$ increases. A stop-motion video of the SEM images is available in the Supplementary Data provided.

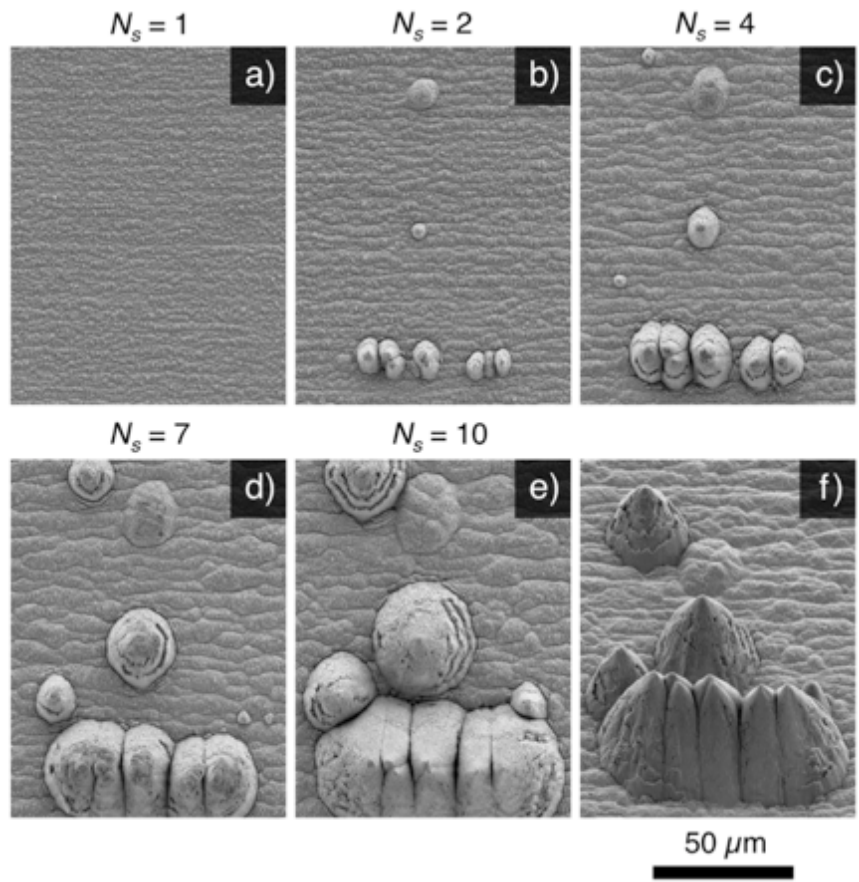

**Figure 3.** Stop-motion SEM images tracking the formation and growth of ellipsoidal cones on stainless steel 304 at $F_{\Sigma line,max} = 47$ J/cm$^2$ and $N_s = 10$. The micrograph in f) was obtained by tilting the sample at an angle of 50º.

When $N_s$ = 1, the surface topography is composed exclusively of LIPSS (Figure 3a). After the second laser scan (Figure 3b), however, round and oblong surface protrusions begin to emerge amidst the LIPSS backdrop. These protrusions, which are essentially small ellipsoidal cones, are randomly distributed over the irradiated area and may exist either as individual bumps or in close proximity to neighbouring ones. These ellipsoidal cones increase steadily in size as they are re-scanned. Taking the ellipsoidal cone located in the centre of each image in Figure 3 as an example, its average diameter increases from 5.5 µm ($N_s$ = 2) to 13.3 µm ($N_s$ = 4), 25.8 µm ($N_s$ = 7), and then finally to 39.1 µm ($N_s$ = 10). In addition, not only do the other ellipsoidal cones present in Figure 3 increase in size as well, but they also merge with neighbouring cones when their bases meet. These observations from the stop-motion SEM images in Figure 3 are clear indications that each ellipsoidal cone originates from a certain surface precursor, which then grows in size as it is rescanned by the fs-laser.

Furthermore, it is evident that the formation of ellipsoidal cone precursors does not solely occur between $N_s$ = 1 and 2. In other words, the surface precursors may appear after any number of laser scans, as demonstrated by Figures 3c ($N_s$ = 4) and 3d ($N_s$ = 7). As a result, the ellipsoidal cones that are found in one single irradiated patch vary in height and diameter since their formation origin begins at different values of $N_s$, and hence at different "times". Finally, while it is clear that ellipsoidal cones originate from surface precursors, not every precursor grows into an ellipsoidal cone. Consider the topmost surface mound found in Figure 3b: as $N_s$ increases, its diameter does not increase at the same rate as the other ellipsoidal cones. The latter are covered in nanoparticles, but the former remains covered by LIPSS. The tilted SEM image in Figure 3f reveals that, at $N_s$ = 10, the "failed" precursor has not taken the shape and morphology of a typical ellipsoidal cone; it has instead merged with the underlying base topography, which is composed of a ribbed terrain covered in orthogonally-oriented LIPSS. This immediately raises the question of why certain surface precursors mature into ellipsoidal cones, while others subside and ultimately vanish as $N_s$ increases. In order to better comprehend the characteristics of the ellipsoidal cones fabricated by fs-laser irradiation, a series of chemical, crystallographic, and topographical analyses were employed.

*3.2. Chemical analysis*

Numerous studies on Kapton® HN polyimide films have shown that, when subject to laser radiation, conical features are produced on the substrate surface [36-40]. These surface cones, first discovered by Dyer *et al.* [37] in 1986, bear a striking resemblance to the ellipsoidal cones discussed in this work and others [11, 19, 20, 35], in terms of morphology, spatial distribution, and trends with respect to experimental conditions. Krajnovich and Vázquez [36] proposed a "radiation hardening" theory which states that, as more laser pulses are applied, the target polyimide surface becomes increasingly enriched in carbon, leading to an elevated ablation threshold. Since stainless steel 304 is an alloy, it is possible that the ellipsoidal cones observed in this work resulted from the presence of different chemical species in the substrate. Direct EDS measurements of the elemental composition of ellipsoidal cones, the underlying LIPSS terrain, and bare polished steel suggest that there is no significant difference in the species present and their mass fractions when compared to one another.

Furthermore, we successfully reproduced ellipsoidal cones on titanium (98.9% purity) at $F_{\Sigma line,max}$ = 47 J/cm$^2$ and $N_s$ = 10, as shown in Figure 4a. EDS analysis of the irradiated titanium sample showed no difference in composition between ellipsoidal cones and the underlying LIPSS-covered terrain – both measurements indicated that all surface features were primarily composed of titanium. The details regarding the EDS analysis performed in this work can be found in Figure S2 of the Supplementary Data. Finally, the nanoparticle-covered pyramids fabricated by Zuhlke *et al.* [19] on nickel substrates corroborate the fact that ellipsoidal cones are not formed due to a compositional dependence. Hence, the radiation hardening model proposed by Krajnovich and Vázquez [36] does not apply to the ellipsoidal cones observed on stainless steel 304 in this work.

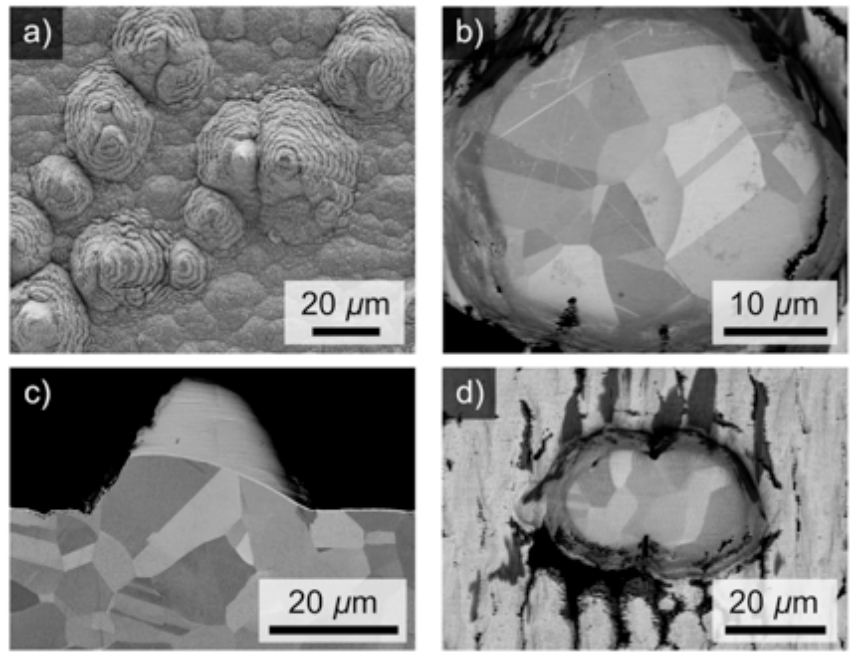

**Figure 4.** a) SEM image of ellipsoidal cones formed on titanium, which was obtained at $F_{\Sigma line,max}$ = 47 J/cm$^2$ and $N_s$ = 10. ECCI images of b) a transverse and c) longitudinal section of two different ellipsoidal cones. d) Transverse section of ellipsoidal cones that merged together. The ECCI images b)-d) of ellipsoidal cones were obtained on stainless steel 304 machined at $F_{\Sigma line,max}$ = 47 J/cm$^2$ and $N_s$ = 10.

*3.3. Crystallographic analysis*

To further probe the formation mechanism of ellipsoidal cones, their crystal microstructure was analyzed by EBSD and ECCI. Ellipsoidal cones at various locations were sectioned and polished as described in Section 2.2.2, and the ECCI images of their transverse and longitudinal cross-sections are shown in Figure 4b-d. The EBSD phase map of the ellipsoidal cone in Figure 4b and others revealed that they are completely comprised of polycrystalline austenite, which is the face-centered cubic (FCC) crystal configuration of stainless steel 304. Therefore, the ellipsoidal cones (excluding their nanoparticle layer) and the base steel substrate exhibit identical phases. In addition, the grain sizes of the ellipsoidal cones are comparable to that of the base stainless steel substrate. The Kikuchi patterns from the EBSD analysis of ellipsoidal cones and the base steel substrate can be found in Figure S3 of the accompanying Supplementary Data.

The ECCI image of the longitudinal cross-section of an ellipsoidal cone in Figure 4c shows that there is, in fact, no interface separating the cone from the base substrate. Instead, grains from the base substrate freely cross into the ellipsoidal cone, which strongly suggests that the substrate and cone are not distinct from one another; rather, the two are essentially one uninterrupted steel structure. Furthermore, when two adjacent ellipsoidal cones merge, as in Figure 4d, the grain boundaries do not follow the contours of the two cones where they merge; instead, the grains extend across the merger point without interruption. Dislocations, which can be detected by ECCI [41], are also largely absent in the crystal structure shown in Figure 4b-d. This is evidence that the steel within the ellipsoidal cones was neither strain hardened nor underwent plastic deformation during fs-laser processing.

Therefore, we conclude that ellipsoidal cones are composed of unablated steel covered by a nanoparticle layer of ablated debris. The similarity in grain sizes between the ellipsoidal cones and steel substrates indicates that during fs-laser ablation, no melting and recrystallization occurs within the cones. Instead, as laser pulses are deposited onto the target, material removal occurs in a layer-by-layer process with each increasing $N_s$, but the ellipsoidal cones stand largely unablated and unaffected by the incoming radiation. Bhattacharya *et al*. [35] and Kam *et al*. [11] propose that, when the fs-laser radiation strikes a surface feature exhibiting sloped sides, the effective fluence sustained is reduced because the incident radiation is spread over a larger surface area relative to a flat plane. Thus, flat surfaces are ablated at a greater rate than sloped structures, a phenomenon known as preferential ablation [18].

In Figure 2, for instance, the nanoparticles on top of the ellipsoidal cone are sintered together while the rest of the cone is covered by loosely organized spherical nanoparticles. At the base of the ellipsoidal cone in Figure 2, its sides are inclined at approximately 62º with respect to the horizontal. The slope begins to quickly level out at the top of the cone, essentially where the nanoparticle layer starts to exhibit a melt-like morphology. Consequently, the ablation rate at the peak of the ellipsoidal cone is greater than at the lower regions. Preferential ablation also explains why not every surface precursor eventually develops into an ellipsoidal cone, which is demonstrated in Figure 3. Inspection of Figure 3f reveals that the degenerated precursor exhibited a very low aspect ratio, which, in turn, resulted in a smaller fluence reduction as compared to the other precursors. As a result, it was quickly ablated when exposed to fs-laser radiation, along with the underlying LIPSS terrain.

*3.4. Aspect ratio and fluence reduction*

The degree of fluence reduction corresponding to cones obtained at different experimental settings was estimated. The heights *h* and diameters *d* of ellipsoidal cones were measured using confocal microscopy for patches irradiated at different values of $F_{\Sigma line,max}$, with $N_s = 10$. Since surface precursors can form at any $N_s > 1$, the ellipsoidal cones found within any given patch can vary greatly in *h* and *d*. Therefore, it is more practical to compute the aspect ratio (*h*/*d*) of the ellipsoidal cones as a function of $F_{\Sigma line,max}$; these results are shown in Figure 5a.

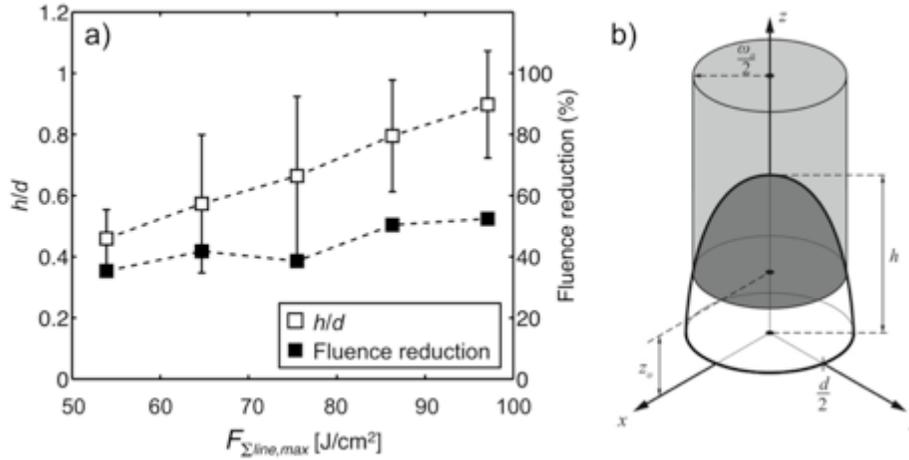

**Figure 5.** a) Aspect (*h*/*d*) ratio of ellipsoidal cones with increasing fluence. The fluence reduction due to preferential ablation is also plotted against $F_{\Sigma line,max}$. The dotted lines are meant to serve as visual guides. b) Schematic of a truncated spheroid used in the estimation of fluence reduction.

In Figure 5a, the mean *h*/*d* ratio increases with $F_{\Sigma line,max}$, which signifies that the sides of the ellipsoidal cones become increasingly inclined at higher pulse energies, regardless of their base diameter. As the aspect ratio of an ellipsoidal cone increases, an incident pulse striking it will be spread over a larger area, which, in turn, leads to a decrease in the effective fluence sustained by the target surface. In this work, the fluence reduction was estimated by modelling each ellipsoidal cone, with height *h* and diameter *d*, as a truncated spheroid, and calculating the total surface area covered by a cylinder with radius equal to that of the beam radius ($\omega_o/2$), as depicted in Figure 5b (The complete derivation of the fluence reduction calculation and the expression for the surface area of a spheroid at different *z* can be found in Section S2 of the Supplementary Data). The approximate fluence reduction calculated based on the mean *h* and *d* for each value of $F_{\Sigma line,max}$ is shown in Figure 5a. The presence of ellipsoidal cones results in a significant decrease in effective fluence: 35% at $F_{\Sigma line,max} = 54$ J/cm$^2$ to 52% at $F_{\Sigma line,max} = 97$ J/cm$^2$. Therefore, if one considers the latter case, flat LIPSS-covered areas are ablated at $F_{\Sigma line,max} = 97$ J/cm$^2$ while the ellipsoidal cones receive approximately half of that fluence level, or $F_{\Sigma line,max} = 48$ J/cm$^2$. Consequently, the flatter terrain is preferentially ablated at a much higher rate than the ellipsoidal cones. In general, the degree of fluence reduction increases with the aspect ratio, and therefore $F_{\Sigma line,max}$. We attribute the slight dip at $F_{\Sigma line,max} = 76$ J/cm$^2$ to the large variance of the measured *h* and *d* values at that particular fluence.

The error bars for the *h*/*d* plot in Figure 5a represent the 95% confidence intervals for each value of $F_{\Sigma line,max}$, measured according to the procedure outlined in Section 2.2.3. While they are undeniably large, there are several important factors that account for the length of the error bars. Firstly, the base of the laser-induced ellipsoidal cones is not exactly circular but is rather an ellipse. Their eccentricities range from 0.40 to 0.75, which indicate that the majority of cones have elliptical bases. Since the length of both semi-major and semi-minor axes were measured for each cone, the measured values of *d* vary appreciably, especially for cones with highly eccentric base areas. Secondly, while using the aspect ratio as an indicator minimizes variations due to the different starting times of ellipsoidal cone growth, it is an imperfect solution because those that begin forming at later $N_s$ are less developed. Finally, not all ellipsoidal cones take the shape of an ellipsoid. Some are deformed due to the fracture and removal of their nanoparticle layers, while others are misshapen and underdeveloped because of their proximity to other taller cones that shield them from incoming laser pulses. This further adds to the variation in the aspect ratios of the ellipsoidal cones. Nevertheless, the preceding topographical analysis provides a reasonably good estimate of the extent of fluence reduction, which, to the best of the authors' knowledge, is presented here for the first time.

*3.5. Formation mechanism*

Figure 6 provides a comprehensive explanation for the formation and growth of the ellipsoidal cones observed in this work. After the stainless steel target is irradiated by fs-laser radiation, material removal by laser ablation causes the original sample surface at $L_0$ to retreat to a new depth $L_1$. The ablated debris, which was ejected from the surface during laser ablation, is deposited and adsorbed onto the newly-created surface (at the ablation depth $L_1$). While most of nano-scale debris is evenly distributed across the irradiated area, some nanoparticles agglomerate together to form larger clusters that collect on the substrate surface. Figure 7 shows evidence for these nanoparticle clusters, whose average size range from 1 to 4.5 μm. Due to the differences in size and morphology of the nanoparticle clusters, only those that provide adequate fluence reduction persist and develop into ellipsoidal cones. We believe that these clusters of nanoparticles, which protrude above the surface level $L_1$, constitute the surface precursors necessary for the formation of ellipsoidal cones. In Figure 6a, two nanoparticle clusters, located at Site A and Site B, serve as precursors to the ellipsoidal cones that develop in Figures 6b and 6c.

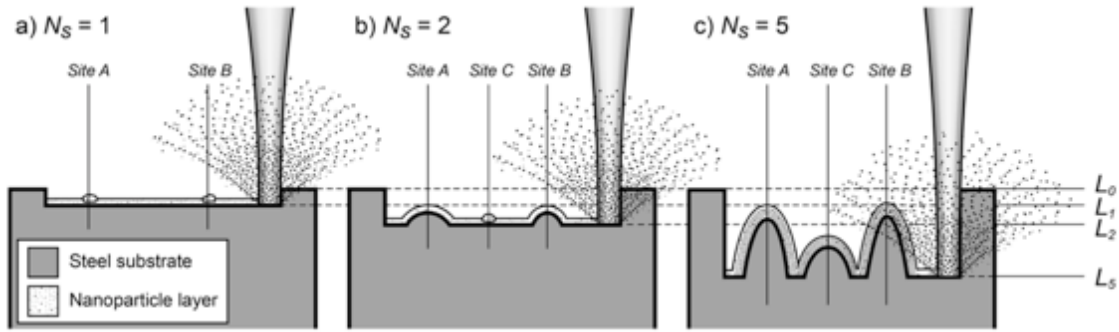

**Figure 6.** Schematic (not drawn to scale) depicting the formation and growth of ellipsoidal cones by multiple fs-laser scanning.

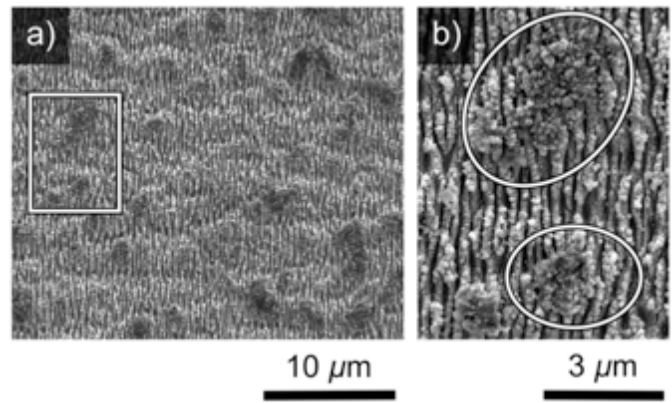

**Figure 7.** a) High magnification SEM image displaying the nanoparticle clusters deposited over the underlying LIPSS topography after one laser scan at $F_{\Sigma line,max} = 45$ J/cm². b) Magnified view of the region indicated in a). The circled regions highlight two nanoparticle clusters for the sake of clarity.

When the irradiated area in Figure 6a is scanned once more (so $N_s = 2$), the target surface is ablated to a new depth $L_2$, which is again covered by redeposited nanoparticulate ablation debris, as depicted in Figure 6b. Due to preferential ablation, the areas located directly beneath the surface precursors at Site A and B in Figure 6a are "shielded" from the incoming radiation and are thus ablated to a lesser degree than the rest of the irradiated area. So, instead of reaching the ablation depth $L_2$ at $N_s = 2$, the steel substrate situated at Site A and Site B remains at a level above $L_2$. Furthermore, due to the ongoing nanoparticle deposition, more randomly distributed surface precursors, in the form of nanoparticle clusters, are generated on the irradiated area, as indicated in Figure 6b at Site C.

Three additional laser passes later ($N_s = 5$, Figure 6c), the substrate surface is ablated to a new depth $L_5$. Material removal from the ellipsoidal cones at Site A and Site B occurs at a much lower rate than the rest of the irradiated area due to the fluence reduction offered by their conical geometry. As a result, while the surface level of the rest of the target area recedes, the substrate contained within the ellipsoidal cone remains intact and thus protrudes above the ablation depth $L_5$. The ellipsoidal cone located at Site C is smaller than those found at Site A and B because its surface precursor only appeared at $N_s = 2$, but the latter two materialized at $N_s = 1$. This means that, while ellipsoidal cones at Site A and Site B began forming from the steel substrate at $L_1$, the cone at Site C only starts at a lower depth of $L_2$. Therefore, ellipsoidal cones that originate at later values of $N_s$ are smaller in size than those that begin forming earlier (see Figure 3).

The ellipsoidal cones in Figure 6c are covered by several distinct layers of nanoparticles, whereby the number of layers corresponds roughly to $N_s$ (as observed in Figures 2 and S1). With each laser pass across the irradiated area, the nanoparticles that are deposited form a single layer atop the previous one. However, the nanoparticles from the previous layer do not fuse with the new one – instead, each layer enveloping the unablated steel interior remains distinctly separate from the next. This is likely due to the fact that, as the previous nanoparticle layer is irradiated by the fs-laser, the nanoparticles are sintered together and thus do not blend with the newly deposited nanoparticles. On the other hand, the nanoparticles covering the flat areas of the irradiated region experience higher fluence and are ablated away during the next laser scan, and the flat terrain is subsequently covered once more by newly-generated nanoparticles.

If the ellipsoidal cones in Figure 6c were scanned further beyond $N_s = 5$, they would eventually merge with each other and share the same base. This was the case for the cones that were sectioned in Figure 4d: since they shared a common base, the grains were free to extend across the merger point of the two cones. Therefore, the fused cones in Figure 4d demonstrate that they did not, strictly speaking, merge, but rather were never separated from the start: as the sample was ablated, the machining plane retreated, exposing more unablated material.

In addition, the trend observed in Figure 5a as to how the aspect ratio of the ellipsoidal cones increases with $F_{\Sigma line,max}$ can be explained using the formation mechanism depicted in Figure 6. At higher pulse energies, only cones with higher aspect ratios survive because the pulse is spread over a larger area, as compared to cones with low aspect ratios. Therefore, a greater degree of fluence reduction is required as $F_{\Sigma line,max}$ increases in order for the ellipsoidal cones to persist. In fact, not only does their aspect ratio increase with $F_{\Sigma line,max}$, but the spatial density of ellipsoidal cones also decreases rapidly with increasing $F_{\Sigma line,max}$, as shown in Figure 8. This is because, at lower fluence, the majority of ellipsoidal cones are able to withstand the incoming fs-laser radiation. As the pulse energy increases, only cones having an aspect ratio high enough are able to persist, while all other cones with lower aspect ratio are eliminated since they do not provide adequate fluence reduction. The surface topography in Figure 8b, in particular, consists of both developed ellipsoidal cones and "flattened" cones that did not survive (such as the degenerate mound in Figure 3). Once the $F_{\Sigma line,max}$ exceeds a certain threshold (approximately 130 J/cm$^2$ according to Figure 8c), the ellipsoidal cones vanish because any surface precursors or existing cones are obliterated by the fs-laser pulses regardless of the degree of fluence reduction offered by their geometry. At this point, the laser-induced surface topography consists only of LIPSS, irrespective of the number of laser passes applied.

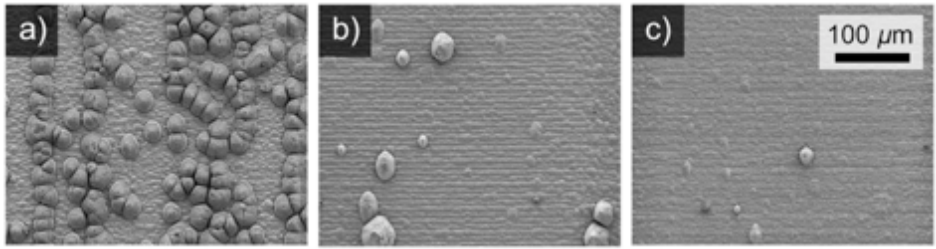

**Figure 8.** Changes in the spatial density of ellipsoidal cones with increasing $F_{\Sigma line,max}$. a) 32 J/cm$^2$, b) 65 J/cm$^2$, c) 129 J/cm$^2$.

Undulating grooves and columnar structures begin to appear as $F_{\Sigma line,max}$ is increased further, which give way to chaotic structures at even higher pulse energies (Section 3). The exact formation mechanism of these laser-induced surface features is still unclear due to the added complexities associated with high energy laser ablation, such as the influence of plasma plume generation and expansion. Nevertheless, it is clear from Section 3 that once the laser-induced surface structures have appeared at $N_s = 1$, the subsequent laser scans do not significantly alter the surface topography (Figure 1). This is in contrast to the ellipsoidal cones observed at much lower $F_{\Sigma line,max}$, which continue to evolve as more laser passes are applied.

## 4. Conclusion

The effect of multiple scanning on the resultant fs-laser-induced surface topographies on stainless steel 304 has been thoroughly investigated in this work. The surface patterns obtained by re-scanning can be separated into two fluence regimes. At low fluence (approximately $F_{\Sigma line,max} < 130$ J/cm$^2$), ellipsoidal cones – composed of unablated steel concealed within several distinct layers of nanoparticles – appear randomly distributed over the irradiated area. In this first regime, deposition dominates over ablation due to the low pulse energies utilized [35]. Nanoparticles that deposit on the machined surface agglomerate and form clusters that serve as precursors for ellipsoidal cones. Due to preferential ablation, material removal from these precursors occurs at a much lower rate than the surrounding flat LIPSS terrain. As a result, these precursors develop into ellipsoidal cones that protrude above the flat LIPSS topography. The second regime begins when the pulse energy is high enough to ablate the surface precursors and ellipsoidal cones despite being spread over a large incident area. Here, ablation dominates over deposition, leading to the formation of other laser-induced surface patterns such as columnar and chaotic structures. These surface topographies are formed immediately at $N_s = 1$, which is in contrast to the formation of ellipsoidal cones, which need at least 2 laser scans to develop. Furthermore, while the columnar and chaotic topographies remain essentially unchanged with $N_s$, the ellipsoidal cones in the first regime spawn and develop continuously as $N_s$ is increased until the entire irradiated area is covered in cones. The investigation of multiple scanning sheds new light upon the formation mechanisms behind fs-laser surface texturing.

## 5. Acknowledgements

The authors gratefully acknowledge the financial support received from the Natural Sciences and Engineering Research Council of Canada (NSERC). We also thank Dr. Samir Mourad Chentouf and Dr. Mohammad Jahazi from l'École de technologie supérieure (ÉTS) for the use of their 3D confocal microscope. Finally, we would like to thank Dr. Nikolas Provatas from the Department of Physics at McGill University for the insightful discussion regarding the possible formation mechanisms of ellipsoidal cones.

# Investigating and understanding the effects of multiple femtosecond laser scans on the surface topography of metallic specimens

Edwin Jee Yang Ling[a], Julien Saïd[a], Nicolas Brodusch[b], Raynald Gauvin[b], Phillip Servio[a], and Anne-Marie Kietzig[a,c]

a. Department of Chemical Engineering, McGill University, 3610 University Street, Montréal, Québec, H3A 0C5, Canada.

b. Department of Mining and Materials Engineering, McGill University, 3610 University Street, Montréal, Québec, H3A 0C5, Canada.

c. Corresponding author. Tel: +1(514) 398-3302. E-mail: anne.kietzig@mcgill.ca

## S1. Supplementary Data

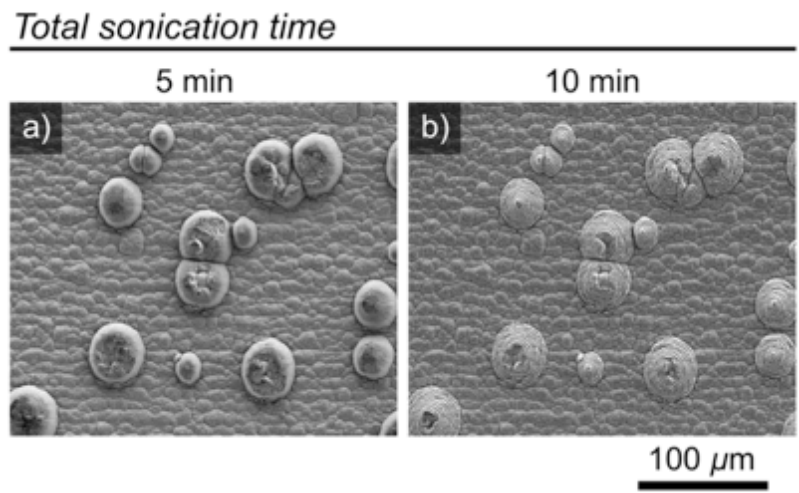

**Figure S1.** SEM images of ellipsoidal cones machined at $F_{\Sigma line,max}$ = 45 J/cm$^2$ and $N_s$ = 10. The stainless steel sample in a) was cleaned with acetone in an ultrasonic bath for 5 minutes prior to scanning electron microscopy (SEM) analysis. b) The same sample was then cleaned for another 5 minutes in the ultrasonic bath with acetone and imaged under the SEM, revealing the underlying nanoparticle strata covering the ellipsoidal cones.

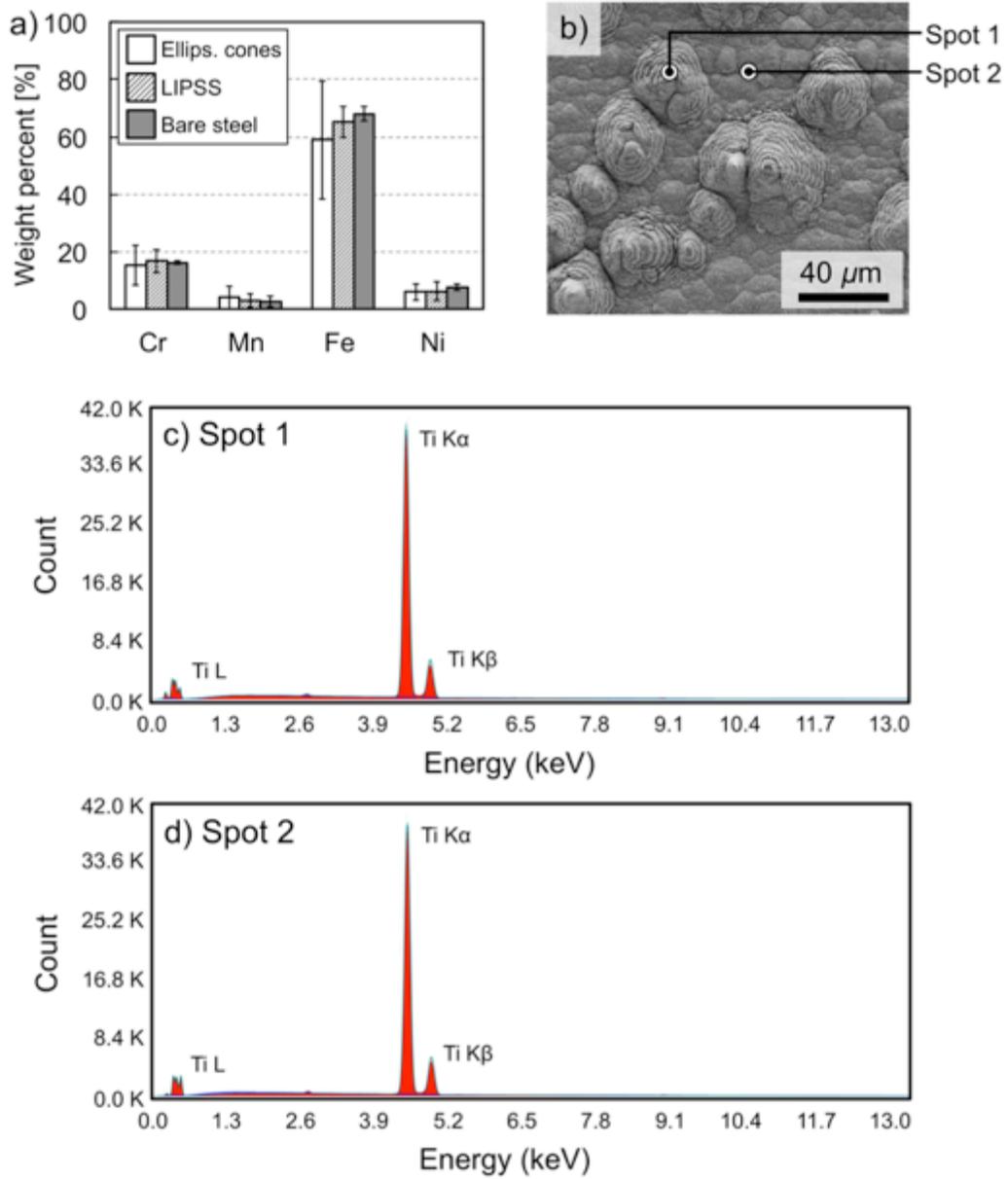

**Figure S2.** a) Mass fraction of different chemical species based on EDS analysis of three different regions on a stainless steel 304 specimen. b) SEM image of ellipsoidal cones produced on titanium at $F_{\Sigma line,max}$ = 47 J/cm$^2$ and $N_s$ = 10. The two spots indicated in the image were analyzed by EDS, and the EDS spectra are displayed in c) and d).

Figure S2a displays the weight percent of chromium (Cr), manganese (Mn), iron (Fe), and nickel (Ni) measured by energy dispersive X-ray spectroscopy (EDS) of three different regions: ellipsoidal cones, the underlying laser-induced periodic surface structures (LIPSS) terrain, and bare stainless steel 304. Both the ellipsoidal cones and the LIPSS terrain were sampled at 40 different locations each, while the bare steel was measured at 20 different locations; the error bars in the bar graph represent 95% confidence intervals of the measured compositions.

The results in Figure S2a indicate that the composition of the main constituents of the three sampled regions are very similar, except for Fe. However, it is difficult to say whether the weight percent of Fe is significantly lower in ellipsoidal cones than in the other two regions due to the large variance in the measured values. The large 95% confidence intervals associated with EDS measurements on ellipsoidal cones can be attributed to two causes: first, the thickness of the nanoparticle layer imparted by one laser pass is approximately 1.4 μm (estimated from SEMs such as that shown in Figure S2), which results in a total nanoparticle layer of at least 13 μm at $N_s$ = 10. Since the interaction depth of the electron beam on the substrate is on the order of 1-2 μm, the electron beam is only exciting electrons within the nanoparticle layer. As seen in Figure 2 of the main article, this nanoparticle layer is, for the most part, made up of clumps of spherical nanoparticles that are arranged in a fragile and porous network. Such an arrangement promotes the absorption of emitted X-rays within the substrate via the photoelectric effect, which results in a lower X-ray count. Furthermore, since the nanoparticles are not chemically homogeneous, variations in the EDS spectra will inherently be present. Second, X-rays emitted from ellipsoidal cones may not be fully received by the silicon drift detector due to the sloped sides of the ellipsoidal cones, which alters the trajectory of the emitted X-rays. Therefore, the composition of species within ellipsoidal cones exhibits a large variance, as demonstrated by the large error bars in Figure S2a.

EDS analysis is unsuitable for elements having an atomic number less than 11 because of the low energy X-rays released by electronic transitions between energy levels. In addition, since the amount of carbon within stainless steel 304 is below the detectability limit of EDS, quantifying the carbon concentration using EDS analysis is unfeasible. Furthermore, the oxygen content could not be reliably measured due to its low concentration and because of peak convolution between the K-peaks of lighter elements and the L-family peaks of Mn, Fe, and Ni. Nevertheless, the EDS measurement of elemental composition in different regions of the target substrate suggest that there is no significant difference in the species present and their mass fractions for ellipsoidal cones, the underlying LIPSS terrain, and bare polished steel.

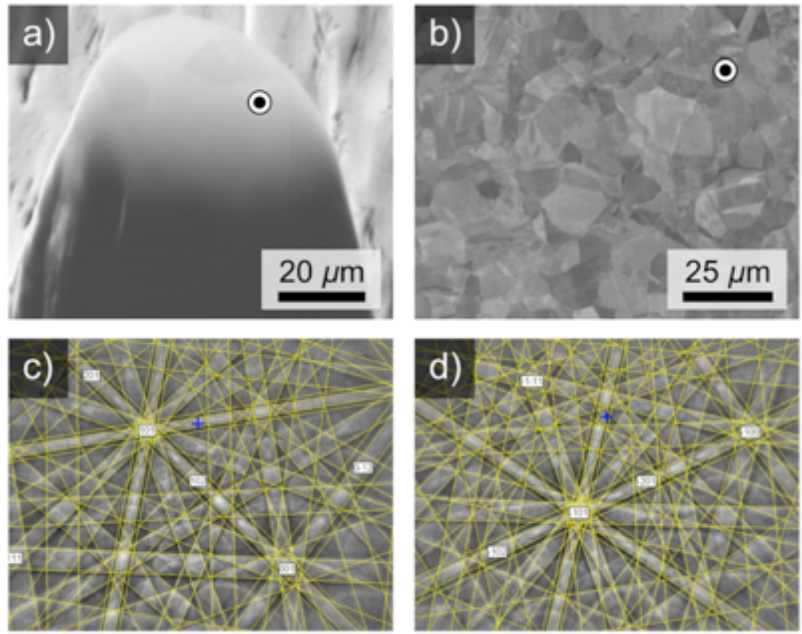

**Figure S3.** SEM images of a) an ellipsoidal cone obtained at $F_{\Sigma line,max}$ = 47 J/cm$^2$ and $N_s$ = 10, and b) polished stainless steel 304 substrate. c) and d) Indexed Kikuchi patterns from the electron backscatter diffraction (EBSD) analyses of the points indicated in a) and b), respectively.

## S2. Calculation of Fluence Reduction of Ellipsoidal Cones

The calculation procedure for the fluence reduction estimation presented in Figure 5a (Section 3.4 of the main article) is shown here. In Cartesian coordinates, the general equation representing an ellipsoid is

$$\frac{x^2}{a^2} + \frac{y^2}{b^2} + \frac{z^2}{c^2} = 1 \tag{S1}$$

In order to simplify the calculation of the surface area $S$ of the ellipsoidal cone, it is assumed that $a = b$, i.e. the intersection of the ellipsoid with the $z = 0$ plane is a circle with a radius of length $a$. Equation S1 then becomes

$$\frac{(x^2 + y^2)}{a^2} + \frac{z^2}{c^2} = 1 \tag{S2}$$

This object is known as a spheroid, which can either be *prolate* ($c > a$) or *oblate* ($c < a$). Re-parameterizing Equation S2 in terms of cylindrical coordinates, we obtain

$$\frac{r^2}{a^2} + \frac{z^2}{c^2} = 1$$

$$r = a\sqrt{1 - \left(\frac{z}{c}\right)^2} \tag{S3}$$

When $\theta = 0$, Equation S3 describes the ellipse found on the $r$-$z$ axes. When this ellipse is rotated about the $z$-axis, the surface of revolution that is generated is the spheroid given by Equation S2. Its surface area $S$ can be calculated as follows

$$S = 2\pi \int r(z)\sqrt{1 + \left(\frac{dr}{dz}\right)^2}\, dz \tag{S4}$$

$$\frac{dr}{dz} = -\frac{az}{c^2\sqrt{1 - \left(\frac{z}{c}\right)^2}} \tag{S5}$$

Substituting Equation S5 and S3 into S4, we obtain the general equation for the surface area of a spheroid

$$S = 2\pi a \int \sqrt{\left[1 - \left(\frac{z}{c}\right)^2\right]\left[1 + \frac{a^2 z^2}{c^4\left(1 - \left(\frac{z}{c}\right)^2\right)}\right]}\, dz$$

$$= 2\pi a \int \sqrt{1 - \left(\frac{z}{c}\right)^2 + \frac{a^2 z^2}{c^4}}\, dz$$

$$= 2\pi a \int \sqrt{1 - \frac{c^2-a^2}{c^4} z^2}\, dz \qquad (S6)$$

Equation S6 is applicable for prolate spheroids ($c > a$). For oblate spheroids ($c < a$),

$$S = 2\pi a \int \sqrt{1 + \frac{a^2-c^2}{c^4} z^2}\, dz \qquad (S7)$$

The derivation that follows will be carried out for prolate spheroids, but the same procedure can be applied to oblate spheroids as well. At this point, it is appropriate to address the limits of integration that should be applied to Equations S6 and S7. The surface area that needs to be computed is not the entire surface area of the ellipsoid, but rather the surface area that is covered by a laser pulse with a diameter $\omega_o$. Assuming that the laser pulse is perfectly circular, its projection onto the surface of the spheroid can be obtained from its intersection with a cylinder of radius $\omega_o/2$, as shown in Figure S4 (reproduced from Figure 5b).

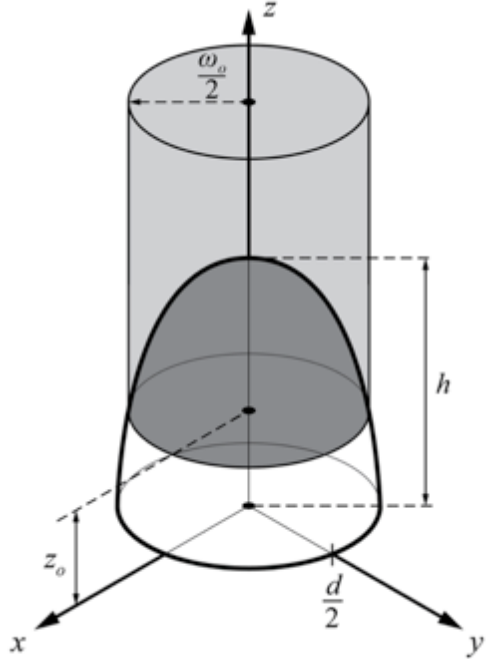

**Figure S4.** Schematic of the intersection of a cylinder and a prolate spheroid. The dark shaded area represents the surface area of the spheroid onto which the cylinder is projected. This area can be calculated by Equations S12 and S13 for a prolate and oblate spheroid, respectively.

The equation of such a cylinder in cylindrical coordinates is

$$r = \frac{\omega_o}{2} \tag{S8}$$

The intersection of the cylinder with the spheroid is obtained by substituting S8 into S3:

$$\frac{\omega_o}{2} = a\sqrt{1 - \left(\frac{z}{c}\right)^2}$$

$$z = c\sqrt{1 - \left(\frac{\omega_o/2}{a}\right)^2}$$

Since $a = d/2$ and $c = h$, where $d$ and $h$ are the diameter and height of an ellipsoid cone, respectively,

$$z_o = h\sqrt{1 - \left(\frac{\omega_o}{d}\right)^2} \tag{S9}$$

where $z_o$ denotes the location of the intersection curve, a circle of radius $\omega_o/2$, on the $z$-axis. Thus, the surface area of a prolate spheroid that is covered by the laser pulse, denoted as $\tilde{S}_P$, is calculated by setting the upper and lower limits of the integral in Equation S6 to $h$ and $z_o$, respectively. Replacing the constants $a$ and $c$, we arrive at

$$\tilde{S}_P = 2\pi a \int_{z_o}^{h} \sqrt{1 - \frac{h^2 - \left(\frac{d}{2}\right)^2}{h^4} z^2}\, dz$$

$$\tilde{S}_P = 2\pi a \int_{z_o}^{h} \sqrt{1 - \beta z^2}\, dz \tag{S10}$$

where

$$\beta = \frac{h^2 - \left(\frac{d}{2}\right)^2}{h^4} \tag{S11}$$

The integral in Equation S10 can be solved by integration by parts, leading to the following expression for $\tilde{S}_P$:

$$\tilde{S}_P = \frac{\pi d}{2}\left\{\left[\frac{\arcsin(\sqrt{\beta}h) - \arcsin(\sqrt{\beta}z_o)}{\sqrt{\beta}}\right] + \left[h\sqrt{1 - \beta h^2} - z_o\sqrt{1-\beta z_o^2}\right]\right\} \tag{S12}$$

where $z_o$ and $\beta$ are given by Equations S9 and S11, respectively. For an oblate spheroid, the expression for the surface area covered by the laser pulse, $\tilde{S}_O$, is:

$$\tilde{S}_O = \frac{\pi d}{2}\left\{\left[\frac{\text{arcsinh}(\sqrt{\gamma}h) - \text{arcsinh}(\sqrt{\gamma}z_o)}{\sqrt{\gamma}}\right] + \left[h\sqrt{1+\gamma h^2} - z_o\sqrt{1+\gamma z_o^2}\right]\right\} \quad (S13)$$

where

$$\gamma = \frac{\left(\frac{d}{2}\right)^2 - h^2}{h^4} \quad (S14)$$

If $\omega_o > d$, the base area of the cylinder in Figure S5 is larger than that of the spheroid. In this case, it is assumed that there are no neighbouring spheroids, and the portion of the cylinder that does not cover the spheroid falls on a flat surface (the plane $z = 0$). Then, the total surface area over which the beam is projected, for a prolate spheroid, is:

$$\tilde{S}_P = \frac{\pi d}{2}\left\{\left[\frac{\arcsin(\sqrt{\beta}h) - \arcsin(\sqrt{\beta}z_o)}{\sqrt{\beta}}\right] + \left[h\sqrt{1-\beta h^2} - z_o\sqrt{1-\beta z_o^2}\right]\right\} + \frac{\pi}{4}(\omega_o^2 - d^2) \quad (S15)$$

Finally, the reduction of fluence due to the spreading of the laser pulse over the ellipsoidal cone can be calculated from the following ratio:

$$\text{Fluence reduction} = 1 - \frac{\tilde{F}_p}{F_o}$$

$$= 1 - \frac{\left(\frac{2P}{\tilde{S}f}\right)}{\left(\frac{2P}{\pi\left(\frac{\omega_o}{2}\right)^2 f}\right)}$$

$$= 1 - \frac{\pi\left(\frac{\omega_o}{2}\right)^2}{\tilde{S}} \quad (S16)$$

where $F_o$ is the peak fluence calculated over a circular area with a diameter $\omega_o$ and $\tilde{F}_p$ is the peak fluence calculated over the area projected by that same circle onto a spheroid (use $\tilde{S}_O$ for an oblate spheroid and $\tilde{S}_P$ for a prolate spheroid in place of $\tilde{S}$), $P$ is the average power of the laser beam, and $f$ is the repetition rate of the laser system.